\documentclass[traditabstract]{aa} % for the abstract without structuration 
\usepackage{txfonts}
\usepackage{natbib}
\bibpunct{(}{)}{;}{a}{}{,}

\usepackage{graphicx}% Include figure files
\usepackage{dcolumn}% Align table columns on decimal point
\usepackage{bm}% bold math
\usepackage{color}

\def\be{\begin{equation}}
\def\ee{\end{equation}}
\def\beqn{\begin{eqnarray}}
\def\eeqn{\end{eqnarray}}

\def\n{{\bf \hat n}}

\newcommand{\wj}[6]{\left(
\begin{array}{ccc}
#1&#2&#3\\
#4&#5&#6
\end{array}
\right)
}

\newcommand{\wsj}[6]{\left\{
\begin{array}{ccc}
#1&#2&#3\\
#4&#5&#6
\end{array}
\right\}
}
\graphicspath{{./plots/}}

\begin{document}

\title{Full-sky CMB lensing reconstruction in presence of sky-cuts}

\author{Aur{\'e}lien Benoit-L{\'e}vy
	\inst{1,2}
	\and
	Typhaine Dechelette\inst{2} \and
	Karim Benabed \inst{2} \and
	Jean-Fran\c{c}ois Cardoso \inst{2,3,4} \and
	Duncan Hanson \inst{5,6} \and
	Simon Prunet\inst{2}
}

\institute{Department of Physics and Astronomy, University College London, London WC1E 6BT, UK
\and Institut d'Astrophysique de Paris, CNRS, UMR7095, Universit{\'e} Pierre\& Marie Curie, 98bis boulevard Arago, 75014 Paris, France
\and Astroparticule et Cosmologie, CNRS UMR7164, Universit{\'e} Denis Diderot Paris 7, B{\^a}timent Condorcet, 10 rue A. Domon et L.~Duquet, 75013 Paris, France
\and Laboratoire Traitement et  de l'Information, CNRS UMR 5141 and T{\'e}l{\'e}com ParisTech, 46 rue Barrault, 75634 Paris Cedex 13, France
\and Jet Propulsion Laboratory, California Institute of Technology, 4800 Oak Grove Drive, Pasadena CA 91109, USA
\and Departement of Physics, McGill University, Montreal QC H3A 2T8, Canada
}
\date{\today}

\abstract{ We consider the  reconstruction of the CMB lensing potential and its power spectrum of the full sphere in presence of sky-cuts due to point sources and Galactic contaminations. Those two effects are treated separately. Small regions contaminated by point sources are filled in using Gaussian constrained realizations. The Galactic plane is simply masked using an apodized mask before lensing reconstruction. This algorithm recovers the power spectrum of the lensing potential with no significant bias.}


\maketitle

\section{Introduction}

The gravitational potential of the large scale structure generates a deflection of the trajectories of the Cosmic Microwave Background (CMB) photons. This effect, known as CMB lensing has long been predicted  (\citealp{Blanchard87, Bernardeau98, Zaldarriaga99}; see  \cite{Lewis06} for a review). It was first detected by cross-correlating CMB data with large-scale structure  \citep{Smith07,Hirata08} and was recently detected internally using only CMB observations \citep{Das11, vanEngelen12}. 
CMB lensing consists in a remapping of the underlying unlensed temperature field. The lensed CMB is related to the unlensed by the deflection angle ${\bf d}$, which is the gradient of the lensing potential (${\bf d}(\n) = \nabla\phi(\n)$). 
The lensing potential $\phi(\n)$ is defined as the integral along the line of sight of the gravitational potential of the large scale structure weighted by a geometrical kernel 
\be
\phi(\n) = -2 \int d\eta \frac{\chi(\eta-\eta_{\rm{rec}})}{\chi(\eta_{\rm{rec}})\chi(\eta)}\Psi(\chi \n, \eta),
\ee
where $\Psi$ is the Newtonian potential, $\eta$ is the conformal time, $\eta_{\rm{rec}}$ is the epoch of last scattering, and $\chi$ is the angular diameter distance in comoving coordinates. The lensing potential thus contains valuable information on the large-scale structure and on the cosmological parameters governing their growth such as the equation of state of Dark Energy or the sum of neutrino masses \citep{Lesgourgues06, abl12b}. It is therefore of utmost importance to be able to recover the lensing potential and its power spectrum from the wealth of high quality CMB data which will soon be available, especially with the forthcoming data of the Planck satellite.

CMB lensing modifies the Gaussian structure of the primary anisotropies and generates a correlation between the temperature and its gradient \citep{Hu00}. These couplings can efficiently be used to construct an estimator, quadratic in the observed temperature, that can be applied to data to recover the lensing potential \citep{Okamoto03}.

Reconstruction of the lensing potential on the full sky requires specially designed techniques to account for the complications inherent to real-data analysis. In particular, the presence of regions of the sky that are too contaminated by Galactic emission or by point sources prevents the direct and straightforward use of the quadratic estimator. Various techniques were developed to account for sky-cuts in a full sky CMB lensing analysis. \citet{Perotto10} developed an algorithm to reconstruct the Gaussian fluctuations of the CMB inside the masked regions. \citet{Smith07} chose to perform the optimal filtering of the observed temperature, by inverting the full variance of the data. This expensive operation, which requires the inversion of a matrix ($N_{pix} \times N_{pix}$) would be feasible at the Planck resolution but requires CPU consuming algorithms. Finally, \citet{Plaszczynski12} recently proposed an hybrid method that decomposes the region outside the Galactic mask in small patches on which a flat-sky analysis can be performed.

Is this article, we present a new and simple method to reconstruct the lensing potential on the sphere. We treat differently the zones of the sky which must be removed because of the presence of point sources and the regions contaminated by the Galaxy.  The pixels corresponding to point sources are inpainted with Gaussian constrained realizations. The idea is to restore isotropic Gaussian CMB fluctuations at the expense of adding a little noise in the data, so that the lensing estimator has the same response as it would have in the ideal case. Regarding the Galactic cut, we choose to apodize the Galactic mask and perform the usual isotropic filtering required by the lensing estimator on the masked data. As we will show later, this operation is almost harmless to the lensing signal.

In Sec. \ref{sec:rec}, we present the formalism used in this paper for lensing reconstruction. Then we treat each aspect separately. In Sec. 3 we investigate the question of the point sources mask and deal with the Galactic mask in Sec. 4. Finally, in Sec. 5 we present a reconstruction where the two effects are present.

\section{Lensing reconstruction}
\label{sec:rec}

We review in this section the formalism of CMB lensing and CMB lensing reconstruction. 
\subsection{Notations}
Throughout the paper, unlensed quantities are denoted with a tilde. Thus, $C_\ell$ denote the lensed TT power spectrum and $\tilde C_\ell$, the unlensed. CMB lensing consists in a remapping of the unlensed temperature on the sky:
\be
\theta[{\bf \hat n}]=\tilde \theta[{\bf \hat n} + {\bf \hat d}], 
\label{eq:cmblens1}
\ee
where $\bf \hat d$ is the deflection angle. This deflection angle can be expressed as the gradient of the lensing potential $\phi$, but for generality can also have a curl component \citep{Hirata03a, Namikawa12}
\be
{\bf \hat d} = \nabla \phi + \nabla \times \psi.
\ee
At first-order approximation, the deflection is a pure gradient and the curl component is null ($\psi=0$), but there exist other mechanisms that could lead to non-zero curl in the deflection angle (e.g., primordial background of gravitational waves from inflation \citep{Cooray05}). Given the smallness of the signal that can be expected in the curl reconstruction we consider in the following that $\psi=0$.

\subsection{Quadratic estimator}

The idea of lensing reconstruction is to rely on the off-diagonal terms in the CMB covariance that are created by CMB lensing. Lensing reconstruction then simply consists in computing the correlations between a filtered version of the observed temperature with a filtered version of its gradient \citep{Okamoto03}. The first step in the reconstruction consists in filtering the observed temperature by its variance. This can be done in several ways, either by inverting the full variance $(S+N)$ \citep{Smith07}, or, simply by applying an isotropic filtering. We chose the latter option and our inverse filtered map then takes the following form
\be
a_{\ell m}= \frac{\theta^{\rm{obs}}_{\ell m}}{C_\ell^{\rm{tot}}},
\ee
where $\theta^{\rm{obs}}_{\ell m}$ are the harmonic multipoles of the observed temperature, and $C_\ell^{\rm{tot}}$ the total power spectrum of the observed temperature. This total spectrum can be decomposed as
\be
C_\ell^{\rm{tot}}= C_\ell^{\rm{fid}} + N_\ell,
\ee
where $C_{\ell}^{\rm{fid}}$ is a fiducial lensed power spectrum and $N_\ell$ the power spectrum of the instrumental noise, which is considered here as white and homogeneous
\be
N_\ell= \Delta T ^2 e^{\ell(\ell+1) \frac{\theta_{\rm{FWHM}}^2}{8\log{2}}}.
\ee
Here $\Delta T$ is the instrumental sensitivity and $\theta_{\rm{FWVM}}$ is the angular resolution parameter of the CMB experiment \citep{Knox95}.

The inverse variance filtered multipoles are then used as a input to the quadratic estimator. We follow the construction of the quadratic estimator by \citet{Okamoto03}, but we slightly modify it along the lines of \citet{Lewis11}. Instead of filtering the inverse variance filtered map by the unlensed temperature power spectrum, we use the lensed power spectrum. This small modification has been shown to greatly simplify the expression of the variance of the lensing estimator \citep{Hanson11} -- hence the estimation of the lensing power spectrum.
To summarize, we consider the following fields
\be
\theta^{(1)}(\n)=\sum_{\ell m}f_1(\ell)\theta^{\rm{obs}}_{\ell m }Y_{\ell m}(\n) , \quad \theta^{(2)}(\n)=\sum_{\ell m}f_2(\ell)\theta^{\rm{obs}}_{\ell m }Y_{\ell m}(\n), 
\ee
with
\be
f_2(\ell) = C_\ell f_1(\ell) =\frac{C_\ell}{C_\ell^{\rm{tot}}}.
\ee

We then consider the product of $\theta^{(1)}$ with the gradient of $\theta^{(2)}$,  $\theta^{(1)}\nabla\theta^{(2)}$ \citep{Hu00}. This quantity is a vector field from which we can extract the curl-free  $\bar{g}_{\ell m}$ and gradient-free $\bar{c}_{\ell m}$ components. 
After calculations, $\bar{g}_{\ell m}$ and $\bar{c}_{\ell m}$ take the following forms:
\be
\bar{g}_{\ell m}= \sum_{\ell_1 \ell_2 m_1 m_2} F(\ell_1, \ell_2, m_1, m_2 ,\ell, m)\theta_{\ell_1 m_1}\theta_{\ell_2 m_2} p^+_{\ell \ell_1 \ell_2 }
\label{eq:QE_grad1}
\ee
and
\be
\bar{c}_{\ell m}= {\bf i}\sum_{\ell_1 \ell_2 m_1 m_2} F(\ell_1, \ell_2, m_1, m_2 ,\ell, m) \theta_{\ell_1 m_1}\theta_{\ell_2 m_2}p^-_{\ell \ell_1 \ell_2 }\ee
where
\beqn
F(\ell_1, \ell_2, m_1, m_2 ,\lambda, \mu)= (-1)^{\mu} f_1(\ell_1) f_2(\ell_2)\sqrt{\ell_2 (\ell_2+1)} \times \\
\sqrt{\frac{(2\ell_1+1)(2\ell_2+1)(2\lambda+1)}{4\pi}}\wj{\ell_1}{\ell_2}{\lambda}{0}{-1}{1}\wj{\ell_1}{\ell_2}{\lambda}{m_1}{m_2}{-\mu},
\eeqn
and 
\be
p^\pm_{\ell \ell1 \ell_2 }= \frac{\left[ 1 \pm(-1)^{\ell+\ell_1+\ell_2}\right] }{2}.
\ee

$\bar{g}_{\ell m}$ and  $\bar{c}_{\ell m}$ are two estimators of the lensing potential and the curl mode.  The expectation of these two estimators are

\begin{equation}
\langle \bar{g}_{\ell m}\rangle =  [A^{\phi}_\ell]^{-1}  \phi_{\ell m} 
\label{meangrad}
\end{equation}

and
\be
\langle \bar{c}_{\ell m}\rangle = 0,
\ee
where $A^{\phi}_\ell$ is a scalar function that renormalizes the estimator so that  the normalized estimator, $g_{\ell m}= A^{\phi}_\ell \bar{g}_{\ell m}$, is unbiased
\be
\langle g_{\ell m}\rangle =   \phi_{\ell m} .
\ee
We can easily show that  $A^{\phi}_\ell$  takes the following form
\beqn
[A^{\phi}_\ell]^{-1}&=& -\frac{1}{4\pi}\sqrt{\ell(\ell+1)}\sum_{\ell_1 \ell_2} f_1(\ell_1) f_2(\ell_2)\\ \nonumber
&& (2\ell_1+1)(2\ell_2+1)\wj{\ell_1}{\ell_2}{\ell}{0}{-1}{1}\sqrt{\ell_2(\ell_2+1)}p^+_{\ell \ell_1 \ell_2} \\ \nonumber
&&\left[ C_{\ell_1}\sqrt{\ell_1(\ell_1+1)}\wj{\ell_2}{\ell_1}{\ell}{0}{-1}{1}  +\right. \\ \nonumber
&&\left.C_{\ell_2}\sqrt{\ell_2(\ell_2+1)}\wj{\ell_1}{\ell_2}{\ell}{0}{-1}{1}     \right] 
\eeqn
Operationally, this quadratic estimator can easily be implemented using the HEALPix  \citep{Gorski2005} routine map2alm\_spin with the two components of $\theta^{(1)}\nabla\theta^{(2)}$ as an input.

It is worth noting that our estimator is different from that of \citet{Okamoto03} by a factor $\sqrt{\ell(\ell+1)}$. This can be explained as \citet{Okamoto03} considered the divergence of  the temperature-gradient product $\theta^{(1)}\nabla\theta^{(2)}$ to extract the curl-free part and then the lensing potential. Their estimator then involves an additional derivative which explain the $\sqrt{\ell(\ell+1)}$ factor.

The variance of the lensing potential estimator is given by \citep{Okamoto03, Kesden03, Hanson11}
\be
\langle g_{LM}g^*_{L^\prime M^\prime}  \rangle= \delta_{LL^\prime}\delta_{MM^\prime}\left[C_L^{\phi\phi}+ N^{(0,g)}_L+ N^{(1,g)}_L\right]
\ee

where $N^{(0,g)}_L$ is a bias term which emerges from the unlensed CMB. More precisely, the $N^{(0)}_L$ bias is the response of the quadratic estimator to the underlying unlensed CMB which has a non-zero unconnected trispectrum \citep{Hu:2001fa}
\beqn
N^{(0,g)}_\ell&=&\frac{[A^{\phi}_\ell]^2}{4\pi} \sum_{\ell_1 \ell_2} f_1(\ell_1) f_2(\ell_2) \sqrt{\ell_2(\ell_2+1)} \\\nonumber
&&(2\ell_1+1)(2\ell_2+1)\wj{\ell_1}{\ell_2}{\ell}{0}{-1}{1}C^{\rm{tot}}_{\ell_1}C^{\rm{tot}}_{\ell_2}p^+_{\ell \ell_1 \ell_2}\\\nonumber
&&\left[  f_1(\ell_1) f_2(\ell_2) \sqrt{\ell_2(\ell_2+1)} \wj{\ell_1}{\ell_2}{\ell}{0}{-1}{1} + \right.\\\nonumber
&&\left.f_1(\ell_2) f_2(\ell_1) \sqrt{\ell_1(\ell_1+1)} \wj{\ell_2}{\ell_1}{\ell}{0}{-1}{1} \right].
\eeqn

The $N^{(1,g)}$ term can be expressed in the full-sky formalism \citep{Hanson11}, but we need to resort to the flat-sky approximation \citep{Kesden03} for a computable expression \citep{Lesgourgues05}.
Following \citet{Hanson11}, $N^{(1,g)}$ takes the following form
\beqn
\label{N1g}
N^{(1,g)}_L&=& \frac{2A^{\phi}_L}{(2L+1)\sqrt{L(L+1)}}\\\nonumber
&&\sum_{\ell_1 \ell_2 \ell_3 \ell_4 L^\prime}(-1)^{\ell_2+\ell_3}\frac{f_{\ell_1 L \ell_2}}{2C^{\rm{tot}}_{\ell_1}C^{\rm{tot}}_{\ell_2} }\frac{f_{\ell_3 L \ell_4}}{2C^{\rm{tot}}_{\ell_3}C^{\rm{tot}}_{\ell_4} }\wsj{\ell_1}{\ell_2}{L}{\ell_4}{\ell_3}{L^\prime}\\\nonumber
&&C^{\phi\phi}_{L^\prime}f_{\ell_1 L^\prime \ell_3}f_{\ell_2 L^\prime \ell_4}p^+_{L \ell_1 \ell_2}p^+_{L \ell_3 \ell_4}p^+_{L^\prime \ell_1 \ell_3}p^+_{L^\prime \ell_2 \ell_4}
\eeqn
with 
\be
f_{\ell_1 L \ell_2}= C_{\ell_2} F_{\ell_1 L \ell_2} +C_{\ell_1} F_{\ell_2 L \ell_1}, 
\ee
and 
\beqn
F_{\ell_1 L \ell_2}= - \sqrt{L(L+1)\ell_2(\ell_2+1)}\wj{\ell_1}{\ell_2}{L}{0}{-1}{1}\\\nonumber
\sqrt{\frac{(2\ell_1+1)(2\ell_2+1)(2L+1)}{4\pi}}.
\eeqn
This expression for $F_{\ell_1 L \ell_2}$ is more general than the one frequently used in the literature and is valid whatever is the parity of the quantity $\ell_1+\ell_2+L$.

We also consider the variance of the estimator of the curl modes. Similarly, we have
\be
\langle c_{LM}c^*_{L^\prime M^\prime}  \rangle= \delta_{LL^\prime}\delta_{MM^\prime}\left[ N^{(0,c)}_L+ N^{(1,c)}_L\right],
\ee
where $ N^{(0,c)}_L+ N^{(1,c)}_L$ are similar to $ N^{(0,g)}_L+ N^{(1,g)}_L$, the only difference being in the parity conditions in their definitions

\beqn
N^{(0,c)}_\ell&=&\frac{[A^{\phi}_\ell]^2}{4\pi} \sum_{\ell_1 \ell_2} f_1(\ell_1) f_2(\ell_2) \sqrt{\ell_2(\ell_2+1)} \\\nonumber
&&(2\ell_1+1)(2\ell_2+1)\wj{\ell_1}{\ell_2}{\ell}{0}{-1}{1}C^{\rm{tot}}_{\ell_1}C^{\rm{tot}}_{\ell_2}p^-_{\ell \ell_1 \ell_2}\\\nonumber
&&\left[  f_1(\ell_1) f_2(\ell_2) \sqrt{\ell_2(\ell_2+1)} \wj{\ell_1}{\ell_2}{\ell}{0}{-1}{1} + \right.\\\nonumber
&&\left.f_1(\ell_2) f_2(\ell_1) \sqrt{\ell_1(\ell_1+1)} \wj{\ell_2}{\ell_1}{\ell}{0}{-1}{1} \right],
\eeqn

\beqn
\label{N1c}
N^{(1,c)}_L&=& \frac{2A^{\phi}_L}{(2L+1)\sqrt{L(L+1)}}\\\nonumber
&&\sum_{\ell_1 \ell_2 \ell_3 \ell_4 L^\prime}(-1)^{\ell_2+\ell_3}\frac{f_{\ell_1 L \ell_2}}{2C^{\rm{tot}}_{\ell_1}C^{\rm{tot}}_{\ell_2} }\frac{f_{\ell_3 L \ell_4}}{2C^{\rm{tot}}_{\ell_3}C^{\rm{tot}}_{\ell_4} }\wsj{\ell_1}{\ell_2}{L}{\ell_4}{\ell_3}{L^\prime}\\\nonumber
&&C^{\phi\phi}_{L^\prime}f_{\ell_1 L^\prime \ell_3}f_{\ell_2 L^\prime \ell_4}p^-_{L \ell_1 \ell_2} p^-_{L \ell_3 \ell_4} p^+_{L^\prime \ell_1 \ell_3}p^+_{L^\prime \ell_2 \ell_4}.
\eeqn

It has been assumed in Eqs. (\ref{N1g}) and (\ref{N1c}) that the power spectrum of the curl modes is null.   It should be noted that the $N^{(1,c)}$ term, despite being related to the curl-like part of the deflection depends on the lensing potential power spectrum. This means than CMB lensing will be a contaminant for any future studies that aim at detecting faint signals in the curl modes of the deflection angle. Here, we follow \citet{Cooray05} and use  the curl  modes in the lensing reconstruction mainly as a systematic test  as we expect a null signal, except for the small $N^{(1,c)}$ term (see also \citet{vanEngelen12} for a flat-sky discussion).

Given the expression of the variance of the quadratic estimator, we can construct an unbiased estimator of the spectrum of the lensing potential \citep{Kesden03}

\be
C_L^{\tilde \phi \tilde\phi}= \langle g_{LM}g^*_{L M}  \rangle -N_L^{(0.g)}- N_L^{(1,g)}.
\label{estgrad}
\ee
This estimator is unbiased and at second order in $\phi$ its  variance is given by
\be
\sigma^2\left(C_L^{\tilde \phi \tilde\phi}\right)=\frac{2}{2L+1}\left[C_\ell^{\phi \phi} +N_L^{(0,g)}+N_L^{(1,g)}\right]^2.
\label{varphi1}
\ee
Similarly, the estimator of the power spectra of the curl modes reads
\be
C_L^{\tilde \psi \tilde\psi}= \langle c_{LM}c^*_{L M}  \rangle -N_L^{(0,c)}- N_L^{(1,c)}.
\label{estcurl}
\ee
The curl term is null at second order in $\phi$ in the expansion of the remapping equation. Besides it can be shown that computations of the next order yield a result compatible with 0. We then have
\be
\langle C_L^{\tilde \psi \tilde\psi}\rangle =0, 
\ee
and the variance is given by
\be
\sigma^2\left( C_L^{\tilde \psi \tilde\psi}\right)= \frac{2}{2L+1}\left[N_L^{(0,c)}+N_L^{(1,c)}\right]^2.
\label{varpsi1}
\ee

\subsection{Simulations}

In order to test the validity of the lensing estimator in the vanilla case of a pure lensed CMB with a perfectly white and homogeneous noise on the full sky, but also in presence of sky-cuts, we generate simulations of lensed temperature field. We use the algorithm presented in \citet{Basak08}. Our lensed temperature maps are then filtered in harmonic space using a Gaussian circular beam  with $\theta_{FWHM}=5\;\rm{arcmin}$ to account for the instrumental beam. We then add to these maps a $\Delta T= 50 \mu\rm{K}.\rm{arcmin} $ homogeneous white noise.  The resulting synthetic maps have characteristics similar to those that the Planck experiment is expected to produce \citep{HFI_DPC}. This map will be pixellized using the HEALPix \citep{Gorski2005} package. Given the experiment characteristics, we will only use the $N_{\rm{side}}=1024$ resolution.

\begin{figure}[htb]
\begin{center}
\includegraphics[width=1.00\columnwidth]{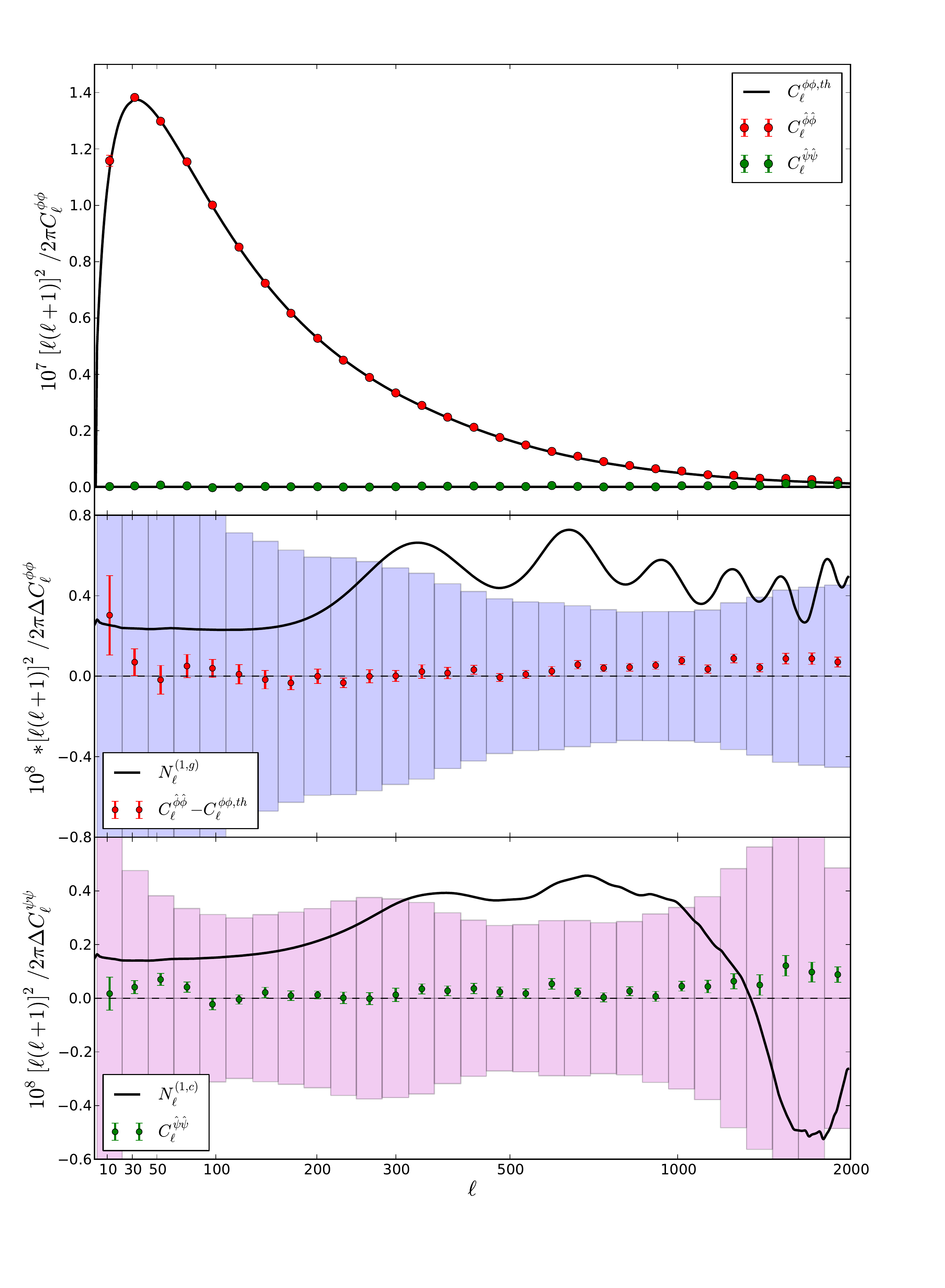}
\caption{\label{fig1} Reconstructed power spectrum of the lensing potential averaged over 300 full-sky lensed simulations. Red dots represent the gradient mode of the deflection (i.e. the lensing potential). Green dots represent the reconstructed curl modes. Middle and bottom panels represent the residuals over 300 lensed simulations of the gradient (middle) and curl (bottom), with the corresponding first-order bias term (black lines). Histogram errors bars represent the theoretical dispersion expected for one single reconstruction (Eqs. \ref{varphi1}) and (\ref{varpsi1}). Error bars on the data points are the realization variance on 300 simulations.}
\end{center}
\end{figure}

As a benchmark of both our implementation of the lensing quadratic estimator and the quality of our simulations, we first try to reconstruct the lensing signal on those perfect full sky synthetic sky realisation. Results are presented in Fig. \ref{fig1}, where we show the average over $N=300$ reconstructions on lensed maps. Thorough examination of the reconstruction for the gradient (middle panel) and the curl (bottom panel) modes reveals the presence of small residuals. Those are small compared to second order corrections ($N_\ell^{(1, gc)}$). They can either come from higher order corrections (that we are ignoring) or lack of convergence of our estimate as we are only using 300 realizations. In the following we will investigate how this residual degrades when taking into account and correct for masks.

Now that we have demonstrated that in the ideal case of an uncut sky the use of the quadratic estimator leads to an unbiased reconstruction, we successively address the questions of  the point source mask and Galactic mask.

\section{Treatment of  the point sources mask}
\begin{figure}[htb]
\begin{center}
\includegraphics[width=1.00\columnwidth]{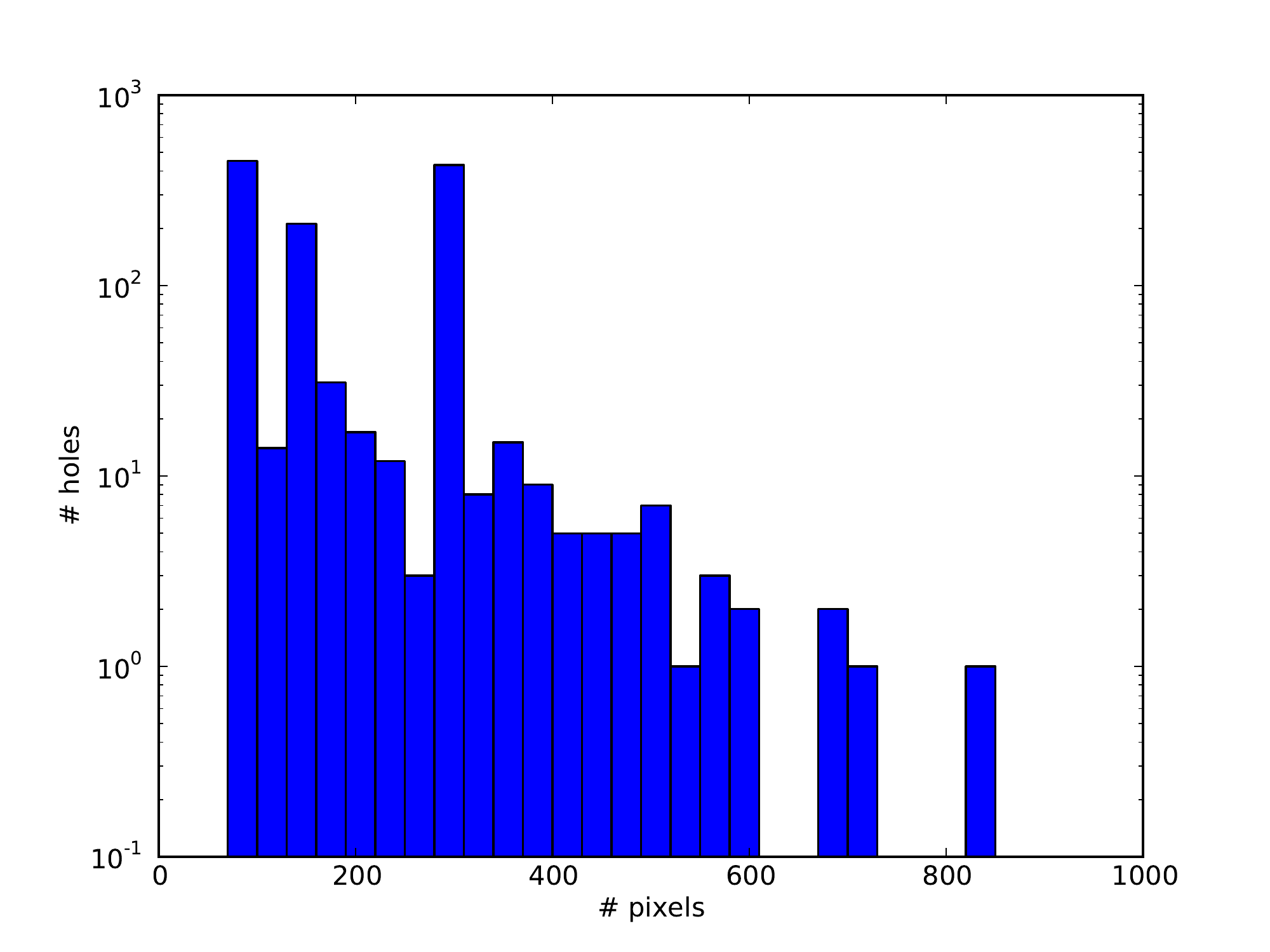}
\caption{\label{hist_ps}Histogram of the sizes of the holes in the point sources mask. }
\end{center}
\end{figure}

In the following results, we use a realistic point source mask constructed from the Planck Early Release Compact Source Catalog \citep{ERCSC2011}. More precisely, we considered the compact sources from the 100, 143 and 217 GHz channels and masked disks with radius equal to three times the values of the beam of the corresponding channel. We therefore use a point sources mask which is composed of holes of various sizes, about 30, 21 and 15 arcmin. Some of the holes overlap, resulting in an enlarged distribution of  holes sizes (Fig. \ref{hist_ps}). We anticipate on the following section, and we will only include in this mask the point sources which are outside a realistic Galactic mask which masks about 20\% of the sky.

\subsection{Filling the point source mask }
\label{sec:cork}

The first step of the lensing reconstruction pipeline consists in restoring a (fake) signal in the region contaminated by resolved point sources. The aim of this operation is to restore the Gaussian statistics of the temperature map to ensure that the response of the quadratic estimator to this inpainted map will be unchanged compared to the full sky case (i.e. without masking). We will do so by  filling the point source holes by a Gaussian realization constrained by the signal around the masked region. This will add a little bit of noise in a small area of the sky ($\approx 2\%$, given the point source mask we use). We will see that this is a small price to be paid for the simplicity gained this way.

We recall here quickly the equation of the Gaussian constrained realizations.  The observed temperature map is decomposed in two parts
\be
T=\left( \begin{array}{c} T_1\\T_2 \end{array} \right ),
\ee
where $T_1$ represents the regions of the sky which are masked and need to be inpainted and $T_2$ the regions of the sky which are used to constrain the value in the masked regions. We also write the covariance matrix  $\Sigma$ of the temperature field as

\be
\Sigma= \left( \begin{array}{cc} \Sigma_{11} & \Sigma_{12}\\ \Sigma_{21} & \Sigma_{22} \end{array} \right) 
\label{eq:corr}
\ee 

The joint probability of $(T_1, T_2)$ is given by
\be
\mathcal{P}(T_1, T_2)\propto \frac{1}{\sqrt{\rm det( \Sigma)}} \exp\left[{-\frac{1}{2}\left( \begin{array}{c} T_1\\T_2 \end{array} \right)^{\dag} \Sigma^{-1} \left( \begin{array}{c} T_1\\T_2 \end{array} \right ) }\right]
\ee

The probability of $T_1$ knowing the constraint $T_2$, is a Gaussian centered on  
\be
\bar{T}_1 =  - W_{11}^{-1} W_{12} T_2 = \Sigma_{12} \Sigma_{22}^{-1} T_2,
\ee
and with a variance
\be
 \sigma= W_{11}^{-1} = \Sigma_{11} - \Sigma_{12} \Sigma_{22}^{-1} \Sigma_{21} 
\ee

Given this formalism, the generation of a local Gaussian constrained realization is straightforward \citep{Hoffman1991}.

\begin{figure}[htb]
\begin{center}
\includegraphics[width=1.00\columnwidth]{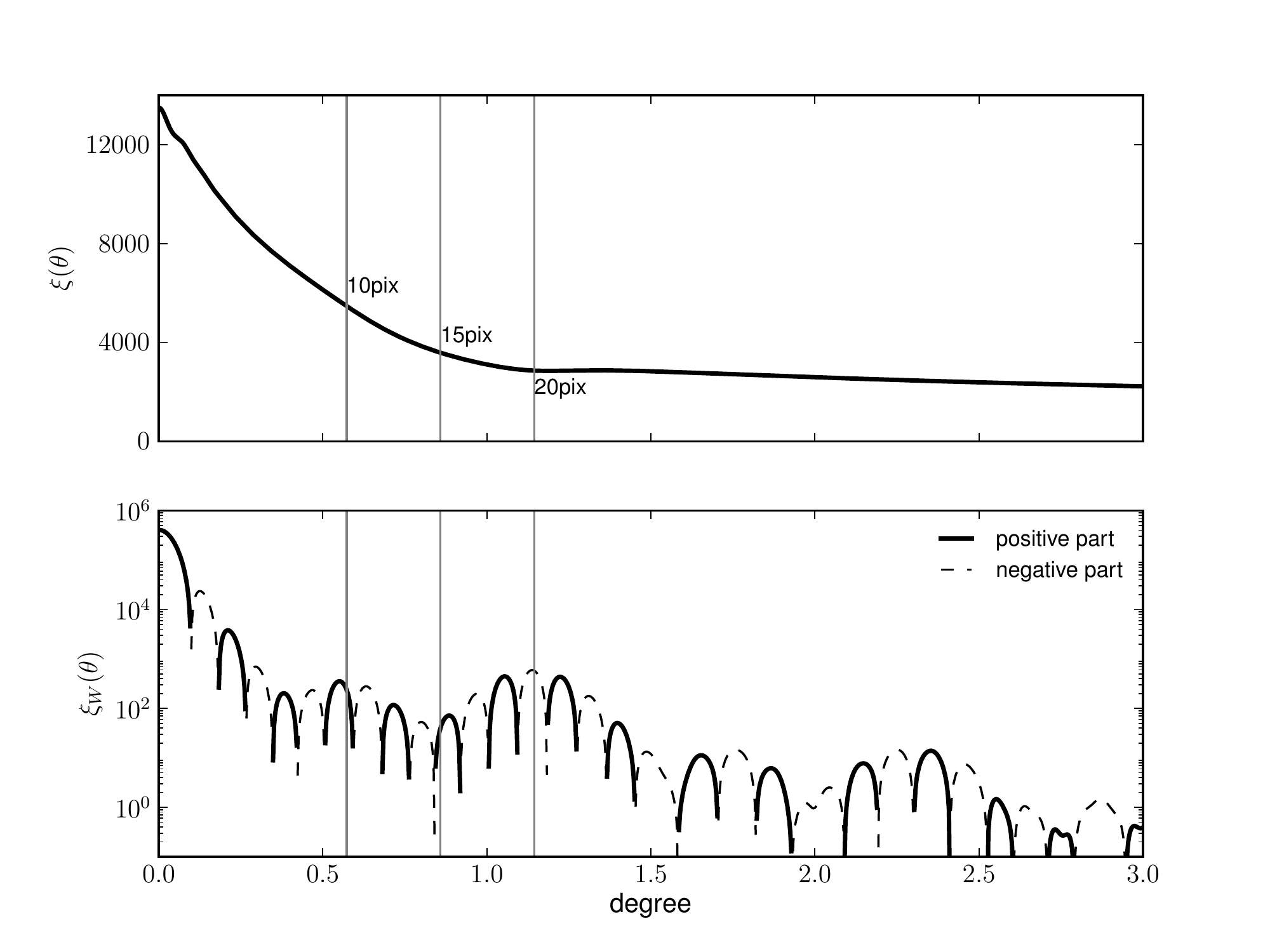}
\caption{\label{constraint} Top panel is the CMB correlation as a function of the angular separation. Bottom, the filter corresponding to $\Sigma_{12}\Sigma_{22}^-1$ as defined in eq. (\ref{corr2}). Vertical lines show the angular separation corresponding to 10, 15 and 20 pixels (i.e. 34.4, 51.5 and 68.7 arcminutes)}
\end{center}
\end{figure}

We start by generating a random realization of the temperature $\tilde T$ with a power spectrum $C_\ell^{\rm{fid}}$.  As the variance of the masked regions given the constraints does not depend on the values of the field, both $T_1$ and $\tilde T_1$ have the same variance. We just need to shift the mean of the inpainted region to recover the expected mean in the masked region. Specifically, we fill the masked regions with the values
\be
T=\tilde{T}_1 + \Sigma_{12}\Sigma_{22}^{-1}(T_2 - \tilde{T}_2). \label{corr2}
\ee
The computation of this quantity requires the inversion of $\Sigma_{22}$. This operation can be prohibitively expensive if we use the full CMB as a constraint. However, given the statistical properties of the CMB temperature anisotropies, the data far away from a given hole will have a small constraint. This is shown in Fig. \ref{constraint} where we represent in real space the CMB correlation function as well as the filter we are constructing above. As we can see the filter is quickly decreasing. This means that we can probably reduce the size of the matrix $\Sigma_{22}$ by only computing it for the unmasked pixel in a narrow region around each hole and get a reasonable approximation of the result. 

\begin{figure}[htb]
\begin{center}
\includegraphics[width=1.00\columnwidth]{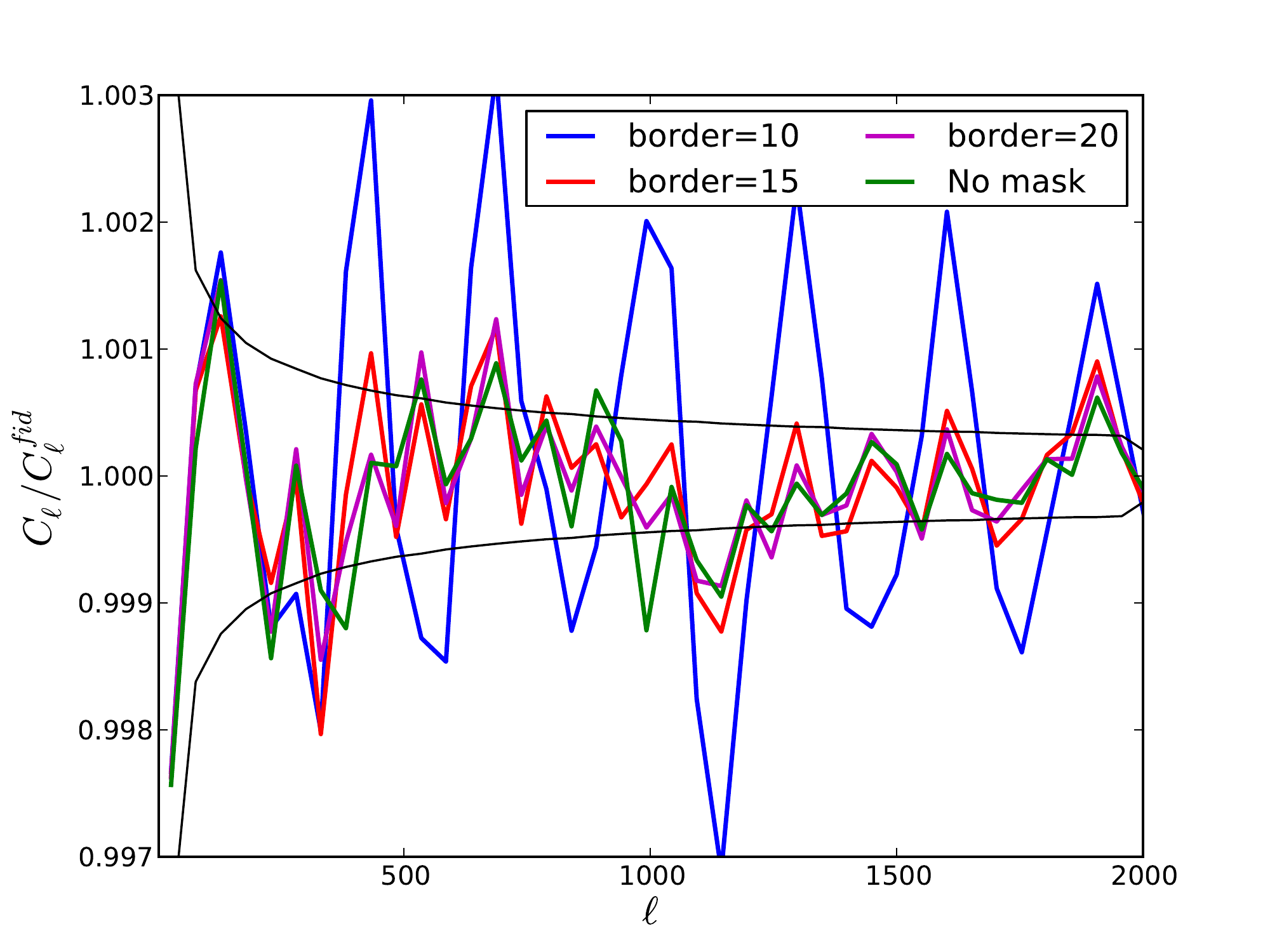}
\caption{\label{border2pt} Ratio of the mean of the N=100 power spectra of the inpainted maps to the fiducial power spectrum for three values of the width of the region used for constraining the Gaussian realization, 10 (blue), 15 (red), and 20 (magenta) pixels. The green line represents the case where no mask is applied. Black lines are the cosmic variance expected given the number of simulations and binning. }
\end{center}
\end{figure}

We perform tests on the Gaussian constrained realizations on our point sources mask with three different choices for the size of the constraint region around the hole, using a 10, 15 or 20 pixel border (resp, 34.4, 51.5 and 68.7 arcminutes with our HEALPix map resolution). The results of these tests are shown in Fig. \ref{border2pt} which presents the ratio between the mean power spectrum of 100 maps whose point source mask as been filled using the procedure described above and the power spectrum used to perform the random realizations. The different results (depending of the size of the border region used to build the constraint) are to be compared with the green line showing the ratio of the mean power spectrum of the unmasked maps to the input power spectrum. This green line, as expected, nicely varies in the region defined by the black lines, which represent the expected variance for 100 realizations in the (slightly wrong at low $\ell$) gaussian approximation for the $C_\ell$.
We clearly see that using only 10 bordering pixels yield to an excess of scatter which is probably correlated given its apparent oscillatory behavior. However using 15 or 20 neighboring pixels seems to be acceptable and at the level of precision of our simulation,  the power spectrum is nicely recovered without any significant bias.

\begin{figure}[htb]
\begin{center}

\includegraphics[width=1.00\columnwidth]{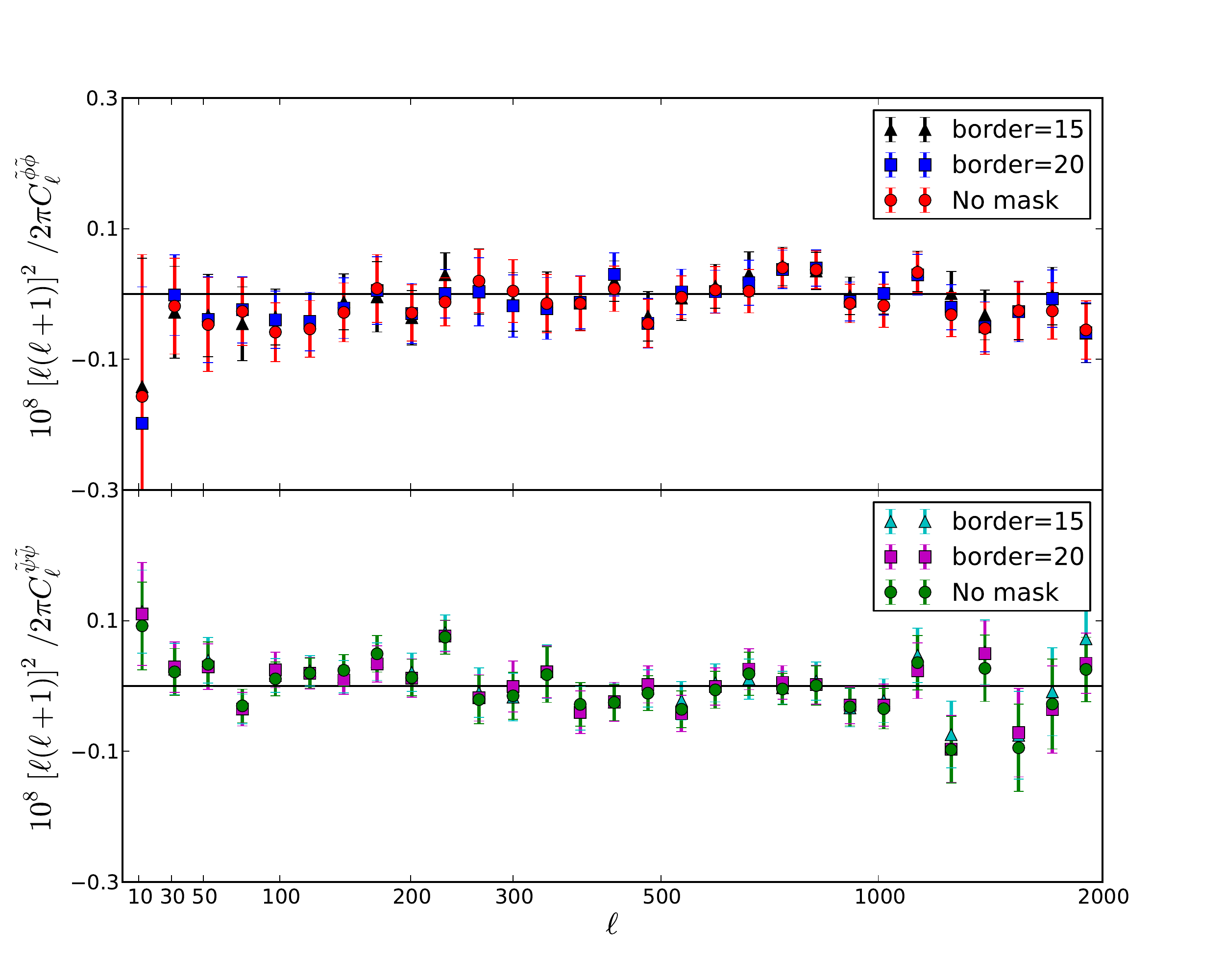}
\caption{ Reconstructed gradient (top) and curl (bottom) modes of the deflection angle on 100 unlensed maps. Point sources are inpainted using 15 (red)  to 20 (magenta) bordering pixels to constrain the Gaussian realizations. The unmasked case  (green) is shown for comparison. \label{rec_ulmap}}
\end{center}
\end{figure}

\subsection{Point sources and lensing reconstruction}

We now wish to demonstrate that our treatment of the point sources mask is harmless for lensing, {\it i.e.} that  1) it does not create any spurious lensing signal, and 2) that it enables an excellent reconstruction of the power spectrum of the lensing potential. We first test the case of unlensed maps. In that case, the terms involving the lensing power spectrum disappear  and according to Eqs. \ref{estgrad} and \ref{estcurl}, the estimated spectrum should be equal to zero.  Here again using around 15 to 20 bordering pixels is required in order to get an unbiased reconstruction. Figure \ref{rec_ulmap} shows the mean over 100 lensing reconstructions on unlensed maps.  Both the cases with 15 and 20 bordering pixels prove to be in very good agreement with the unmasked case, for both the gradient and curl modes. We thus confirm that the point source inpainting does not create any artificial signal that could be seen as related to lensing via the lensing estimator. 

We now turn our analysis on lensed maps. In that case, we need to modify the estimator of the lensing power spectrum to correct for the fact that the regions that are inpainted are filled with a perfectly Gaussian realizations and should therefore not contain any lensing signal. The outputs of the lensing estimator must therefore by rescaled by a factor that corresponds to the fraction of the signal that is masked. This factor should in principle be scale dependent as the large scale of the lensing potential are hardly affected by masking sub-degree scale pixels. However we found that using a scalar scale-independent renormalization gives an excellent reconstruction of the lensing power spectrum. The modified estimators thus reads

\be
C_L^{\tilde \phi \tilde\phi, PS}= \frac{\langle g_{LM}g^*_{L M}  \rangle -N_L^{(0,g)}}{f_{\rm{PS}}^2}- N_L^{(1,g)},
\label{estgradPS}
\ee
where  $f_{\rm{PS}}$ correspond to the fraction of the sky which is unmasked by point sources. With our fiducial point source mask, $f_{\rm{PS}}=0.98$.

\begin{figure}[htb]
\begin{center}
\includegraphics[width=1.00\columnwidth]{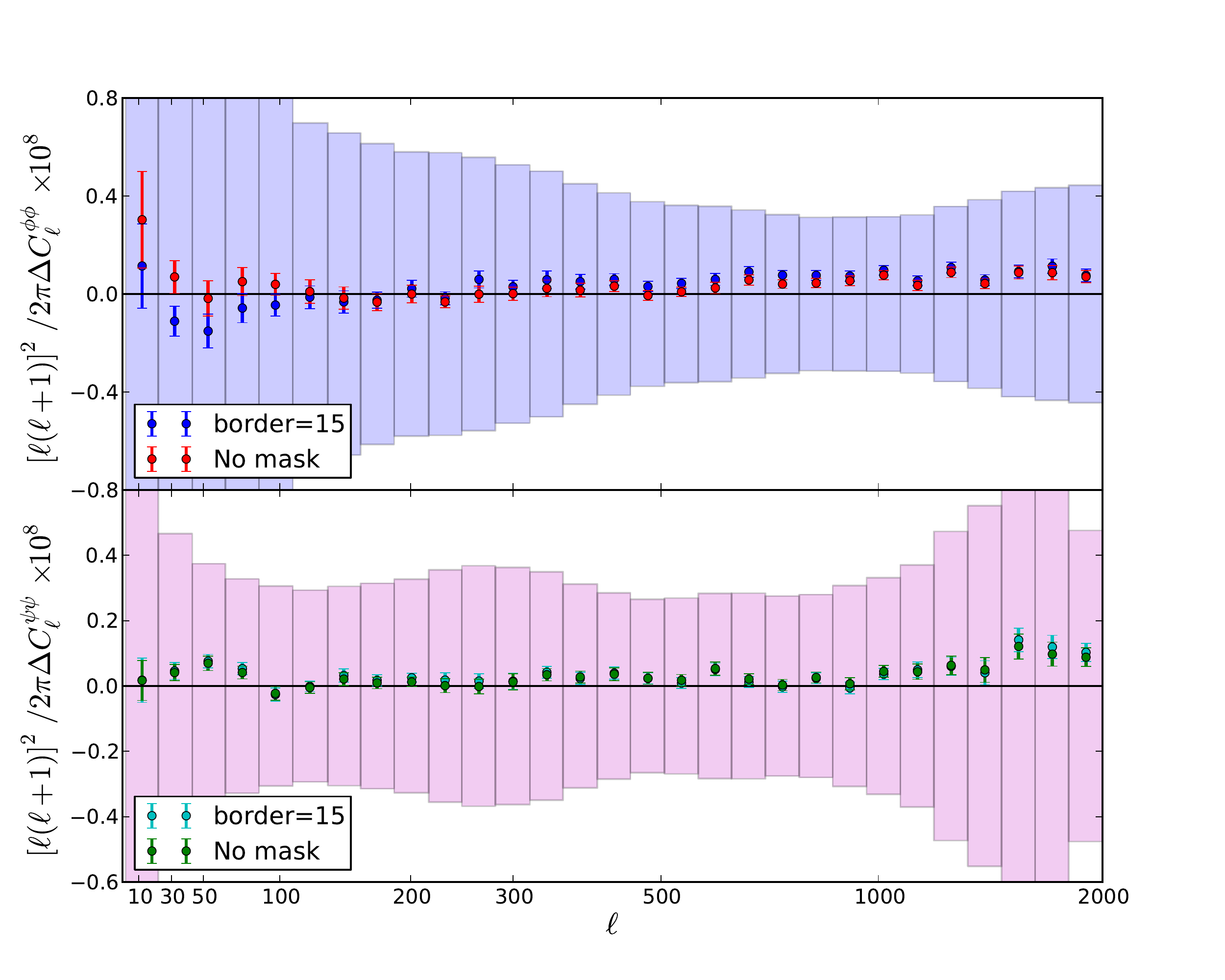}

\caption{Residuals of the reconstructed power spectrum of the lensing potential averaged over 300 full-sky lensed simulations in presence of the point sources mask. Top panel presents the gradient mode of the deflection (i.e. the lensing potential). Bottom panel presents the reconstructed curl modes. In both cases, we compare with the full-sky case (red for the gradient modes and green for the curl modes) Histogram errors bars represent the theoretical dispersion expected for one single reconstruction.  \label{res_cork_lmap}}
\end{center}
\end{figure}

Figure \ref{res_cork_lmap} represents the residuals of the reconstruction, {\it i.e.} \mbox{$\Delta C_\ell^{\phi\phi}=C_\ell^{\tilde \phi \tilde\phi, PS}- C_\ell^{\phi\phi, {\rm{fid}}}$} averaged over 300 simulations.   We also show the residuals of the curl estimator. We compare those results to the full-sky case, with no inpainting. As can be seen, there is no significant difference between the two cases, which validates the use of our inpainting algorithm for lensing reconstruction.

We therefore have a tool that can be efficiently used to inpaint the point sources and that can recover both the 2-point and  4-point statistics of a Gaussian field without introducing any spurious signal or bias. 

\section{Treatment of the Galactic mask}

The second problem frequently encountered in full-sky lensing reconstruction is the presence of large-scale foregrounds, mainly along the Galactic plane, that strongly contaminates the lensing signal. As mentioned in the introduction, several possibilities exist to deal with this complication. We choose a new approach which consists in masking the contaminated regions of the Galactic plane and then directly applying the lensing estimator on those masks maps. We first begin by presenting some analytical computations on the effect of the mask on the reconstructed lensing potential, and we show how these effects can be efficiently alleviated to provide a robust and rapid way of reconstructing the lensing potential.

\subsection{Analytical computation of the mask}

In presence of a mask, the expectancy of the lensing estimator become

\be
\langle g_{LM} \rangle = \sum_{\lambda \mu} M_{LM}^{\lambda \mu} \phi_{\lambda \mu} + \bar \phi_{LM}^{\rm{MF}},
\ee
where  $M_{LM}^{\lambda \mu}$  is a coupling matrix which replaces the scalar normalization $A_L^{\phi}$, and  $\bar \phi_{LM}^{\rm{MF}}$ is a lensing independent quantity which we will loosely refer to as the ``†mask mean field". 

The variance of the estimator applied on a masked temperature map becomes an intricate expression
\be
\langle g_{LM} g^*_{L M}\rangle = \sum_\lambda C_\lambda^{\phi\phi} M^{\lambda}_{LM} + N^{(0,g)}_{L,M}+ N^{(1,g)}_{L,M}+ C_L^{\rm{MF}}, 
\label{eqmaskcl} \ee
where $C_L^{\rm{MF}}$ denotes the power spectrum of the mask mean-field. $N^{(0,g)}_{L,M}$ and  $N^{(1,g)}_{L,M}$ are terms similar to the full-sky equivalent $N^{(0,g)}_{L}$ and  $N^{(1,g)}_{L}$ but also depends on the structure of the mask.
All the terms in the two previous equations involve high order integrals of functions which can be written as intricate summations of Wigner 3-j symbols. They are directly related to the structure of the mask and depend on its harmonic coefficients $w_{\ell m}$. Those functions and integrals can be written down analytically, but cannot lead to a form which can be efficiently evaluated numerically. 
However, we find that under certain circumstances, notably  depending on the mask (and we will provide precise example in Sec. \ref{sec:apo}), those coupling matrices are essentially diagonal, in the sense that their effect on power spectra is just to apply a constant normalization factor. This normalization factor is related to the fraction of the sky which is unmasked \citep{Hivon02}. More precisely, we have
\be
 \sum_\lambda C_\lambda^{\phi\phi} M^{\lambda}_{LM} \approx f_{\rm{gal},4} C_L^{\phi\phi}
 \ee
 and 
 \be
  N^{(0,g)}_{L,M}\approx  f_{\rm{gal},4} N^{(0,g)}_{L}, \; N^{(1,g)}_{L,M}\approx  f_{\rm{gal},4} N^{(1,g)}_{L}, 
 \ee
where 
\be 
f_{\rm{gal,4}}=\frac{1}{N_{\rm{pix}}} \sum_i w_i^4.
\ee
The presence of the coefficients of the mask at the power of four is related to the fact that the variance of the lensing estimator is a resummation of the 4-point correlation function on the CMB.  
The variance of the lensing estimator then becomes
\be
\langle g_{LM} g^*_{L^\prime M^\prime}\rangle = \delta_{LL^\prime}\delta_{MM^\prime}f_{\rm{gal},4}\left[ C^{\phi\phi}_{L} + N^{(0,g)}_{L}+ N^{(1,g)}_{L}\right]+ C_L^{\rm{MF}}, 
\ee

Even though all these mask kernels cannot be computed, some of them are fairly easy to estimate by a Monte-Carlo procedure. Applying the estimator on unlensed maps which have the same spectral content as the lensed map will precisely give the mask mean field
\be
\langle g^{\rm{unl}}_{LM} \rangle = \bar \phi_{LM}^{\rm{MF}},
\ee
It is thus straightforward to construct an unbiased estimator of $\bar \phi_{LM}^{\rm{MF}}$ by averaging over several outputs of the estimator applied on unlensed masked maps
\be
\hat{\phi}^{\rm {MF}}_{LM}= \frac{1}{N_{\rm unl}^{\rm MF}} \sum _i^{N_{\rm unl}^{\rm MF}} g_{LM,i}^{\rm{unl}},
\ee
 
$\bar{\phi}^{\rm {MF}}$ is an unbiased estimate of the mask mean-field and its variance can be approximated by
\be
\langle\hat{\phi}^{\rm {MF}}_{LM}  \hat{\phi}^{\rm {MF *}}_{{L^\prime M^\prime}} \rangle= \delta_{LL^\prime}\delta_{MM^\prime}\left( C_L^{\rm{MF,mask}} + \frac{f_{\rm{gal},4}}{N_{\rm unl}^{\rm MF}} N^{(0,g)}_L\right).
\ee

We can then define a mask mean-field debiased estimator by subtracting the estimated mask mean-field to the lensing potential reconstructed from the masked temperature field. We then define
\be
\hat{g}_{LM}^{\rm{MF}}= g_{LM}- \hat{\phi}^{\rm {MF}}_{LM}.
\ee
The mean of this estimator is
\be
\langle \hat{g}_{LM}^{\rm{MF}} \rangle = \sum_{\lambda \mu} M_{LM}^{\lambda \mu} \phi_{\lambda \mu} ,
\ee
and its variance becomes
\be
\langle \hat{g}^{\rm{MF}}_{LM} \hat{g}^{\rm{MF}*}_{L^\prime M^\prime }\rangle \approx  \delta_{LL^\prime}\delta_{MM^\prime}  f_{\rm{gal},4}\left[ C_L^{\phi\phi} + \left(1+\frac{1}{N_{\rm unl}^{\rm MF}} \right)N^{(0,g)}_{L} + N^{(1,g)}_{L}\right], 
\ee
The removal of the mask mean-field by Monte-Carlo induces a small increase in the variance of the estimator, but this increase can be as small as required by averaging over a large number of unlensed realizations, which is a numerically cheap operation as it does not require the production of lensed simulations.
 
The estimator of the lensing power spectrum now takes the following form
\be
C_L^{\tilde \phi \tilde\phi, mask}= \frac{\langle \hat{g}^{\rm{MF}}_{LM}\hat{g}^{\rm{MF*}}_{L M}  \rangle }{f_{\rm{gal},4}}- \left(1+\frac{1}{N_{\rm unl}^{\rm MF}} \right)N_L^{(0.g)}- N_L^{(1,g)}\label{rec_clpp_mask},
\ee

It should also be noted that on the curl estimator side, the situation is little simpler as there is no mask-mean field (Fig. \ref{rec_mask}, bottom panel). Therefore there is no need  to subtract a "curl" mask mean field to the curl estimate.

If we now consider the variance of the estimator of the lensing power spectrum, it will depend on 
\be 
f_{\rm{gal,8}}=\frac{1}{N_{\rm{pix}}} \sum_i w_i^8,
\ee
as it involves the 8-point correlation function \citep{Kesden03, Hanson11}. We will then have
\be
\sigma^2\left(C_\ell^{ \tilde \phi \tilde \phi, \rm{mask}}\right)= \frac{f_{\rm{gal},8}}{f^2_{\rm{gal},4}}\frac{2}{2\ell+1}\left[ C_\ell^{\phi\phi, \rm{fid}} + N^{(0,g)}_\ell + N^{(1,g)}_\ell   \right]^2
\label{var_mask}\ee

We have set up the formalism to perform a reconstruction on a masked field. We will now give precise examples and explain the regime of validity of our hypothesis.

\subsection{Apodization}
\label{sec:apo}
\begin{figure}[htb]
\begin{center}
\includegraphics[width=1.00\columnwidth]{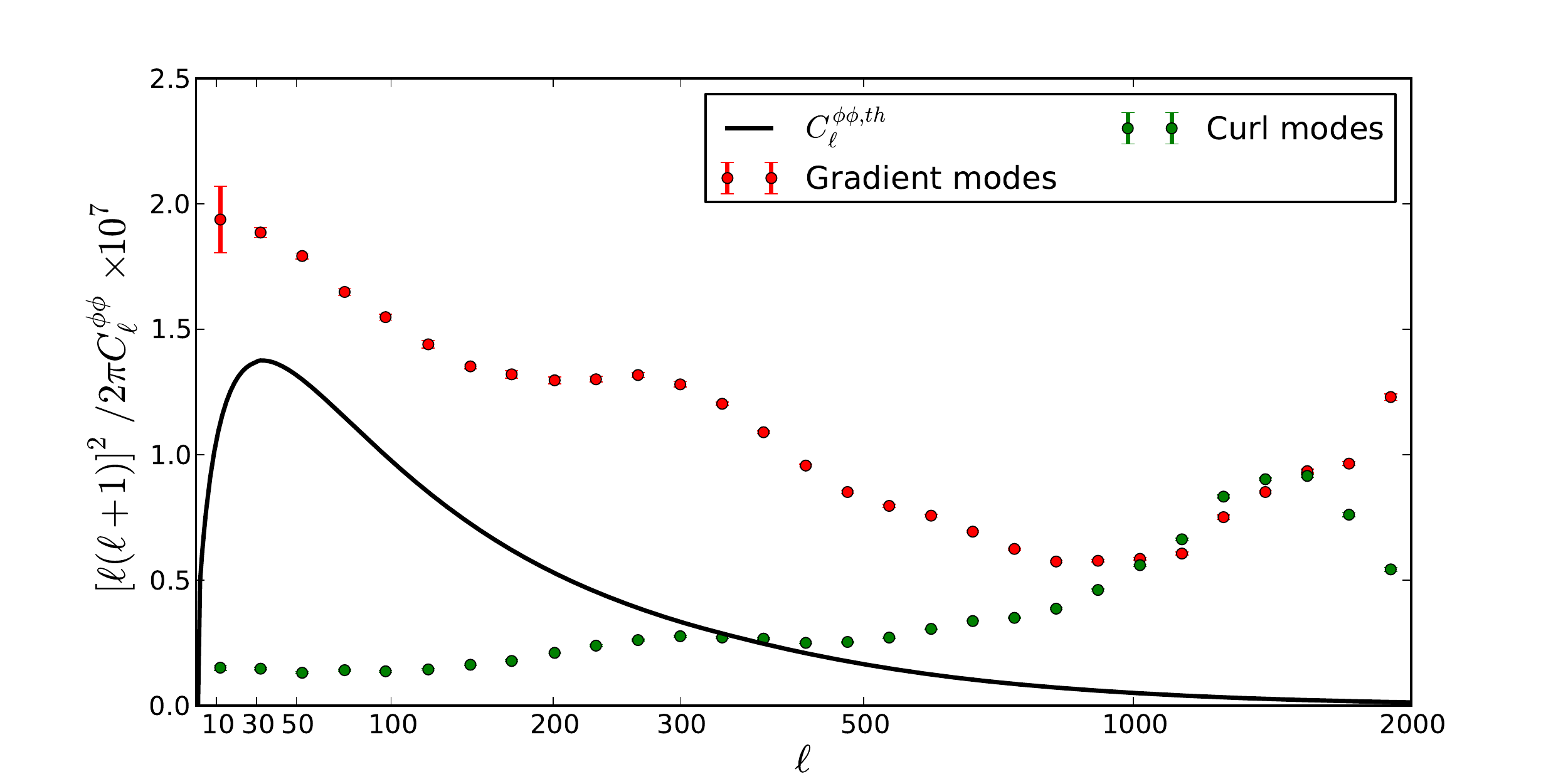}
\caption{Reconstructed lensing potential (Eq. \ref{rec_clpp_mask}) over 100 lensed simulations using the binary mask (no apodization). Green dots are the gradient mode which should overlap the black line, and magenta dots the curl modes, which should be compatible with zero.\label{fig_noapo}}
\end{center}
\end{figure}

If we apply the quadratic estimator directly to the masked temperature fields, both the mask mean field and the band couplings caused by the presence of the mask will highly bias the reconstruction (Fig. \ref{fig_noapo}), both for the gradient and curl estimator. This pathologic behavior is caused by the sharp transition at the boundary of the mask, that creates important power leakage at all scales.

In order to reduce the off-diagonal terms in the coupling matrix and the frequency extent of the mask mean field, we simply need to first apodize the mask before masking the temperature field. Apodization smoothes the spectral content of the mask and reduces the mask-mean field amplitude and removes almost all the off-diagonal terms. This operation has nevertheless a drawback as it decreases the quantity of data used for reconstruction and therefore degrades the statistical significance of the reconstruction. The apodization of the mask is performed by applying a cos-like function to the pixels bordering the mask so that the apodized region goes smoothly from 0 to 1 over the apodization length $\theta_{\rm{apo}}$. 

\begin{figure}[htb]
\begin{center}
\includegraphics[width=1.00\columnwidth]{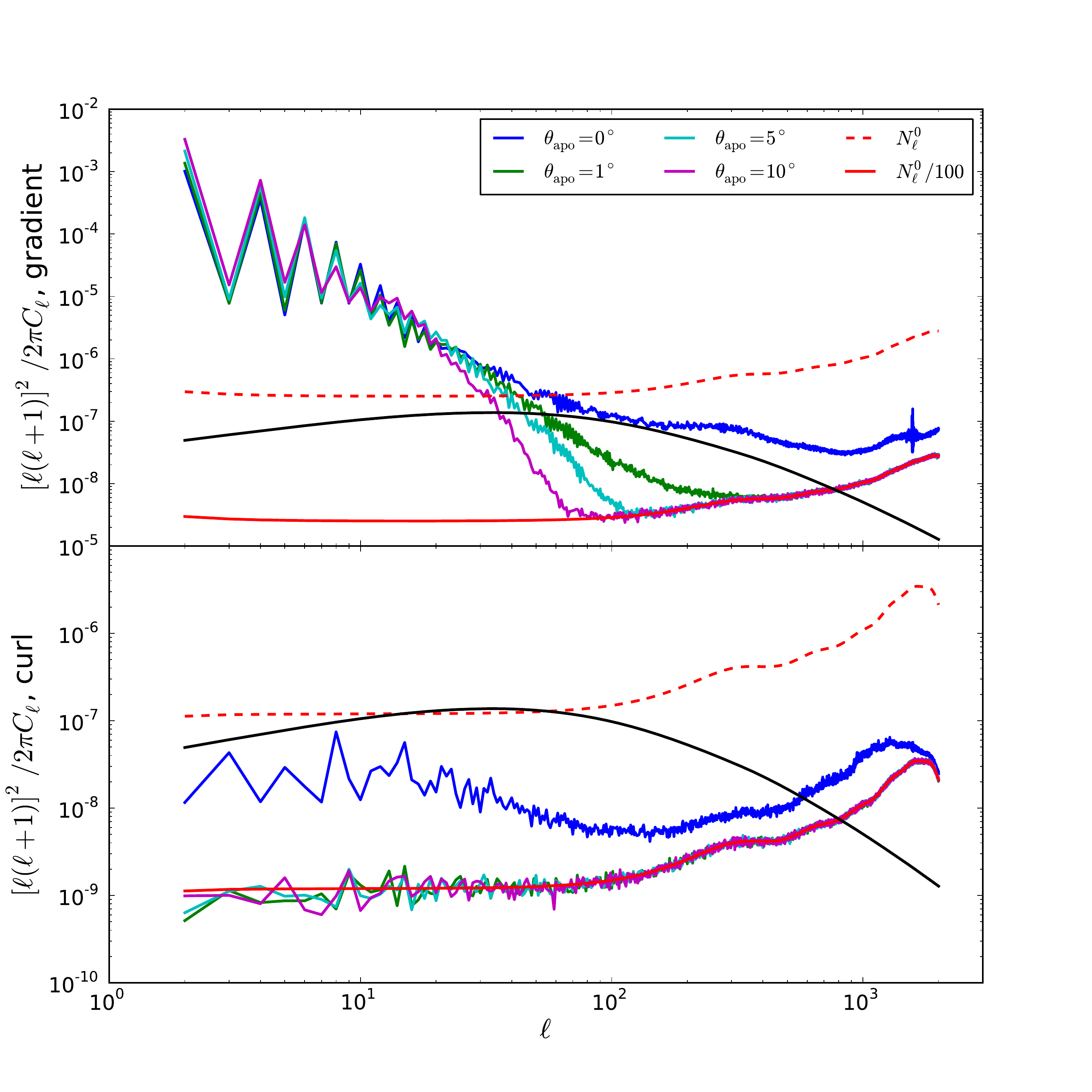}
\caption{Power spectrum of the mask mean field rescaled by $f_{\rm{gal}}$\label{MF_gal} (top, gradient modes; bottom curl modes). The full red line is the Gaussian noise $N^{0, gc}$ divided by the number on unlensed simulation, N=100. When the binary Galactic mask is applied (blue lines) the reconstruction is strongly biased at high multipoles. Note the absence of a mask mean-field in the curl modes.}
\end{center}
\end{figure}

Figure \ref{MF_gal} presents the variance of the estimator of the mask mean field (or equivalently the estimated power spectrum of the mask mean-field) for both the gradient (top panel) and curl (bottom panel) estimators and for different values of the apodization length. In all cases, the variance has been rescaled by the corresponding $f_{\rm{gal},4}$ factor. In this figure, $N_{\rm{unl}}^{\rm{MF}}=100$. We here confirm that the mask does not create a mean-field in the curl estimator (bottom panel). We also see that even a small apodization over 1~degree is enough to remove all the off-diagonal terms and restore the ability of the quadratic estimator to successfully reconstruct the lensing power spectrum.
On the top panel, we clealy see that the mean-field frequency extent decreases when the apodization length is increased. We therefore choose the higher value, $\theta_{\rm{apo}}=10^\circ$ to perform the reconstruction. With this value of apodization length, we have $f_{\rm{gal},4}=0.65$.

\begin{figure}[htb]
\begin{center}
\includegraphics[width=1.00\columnwidth]{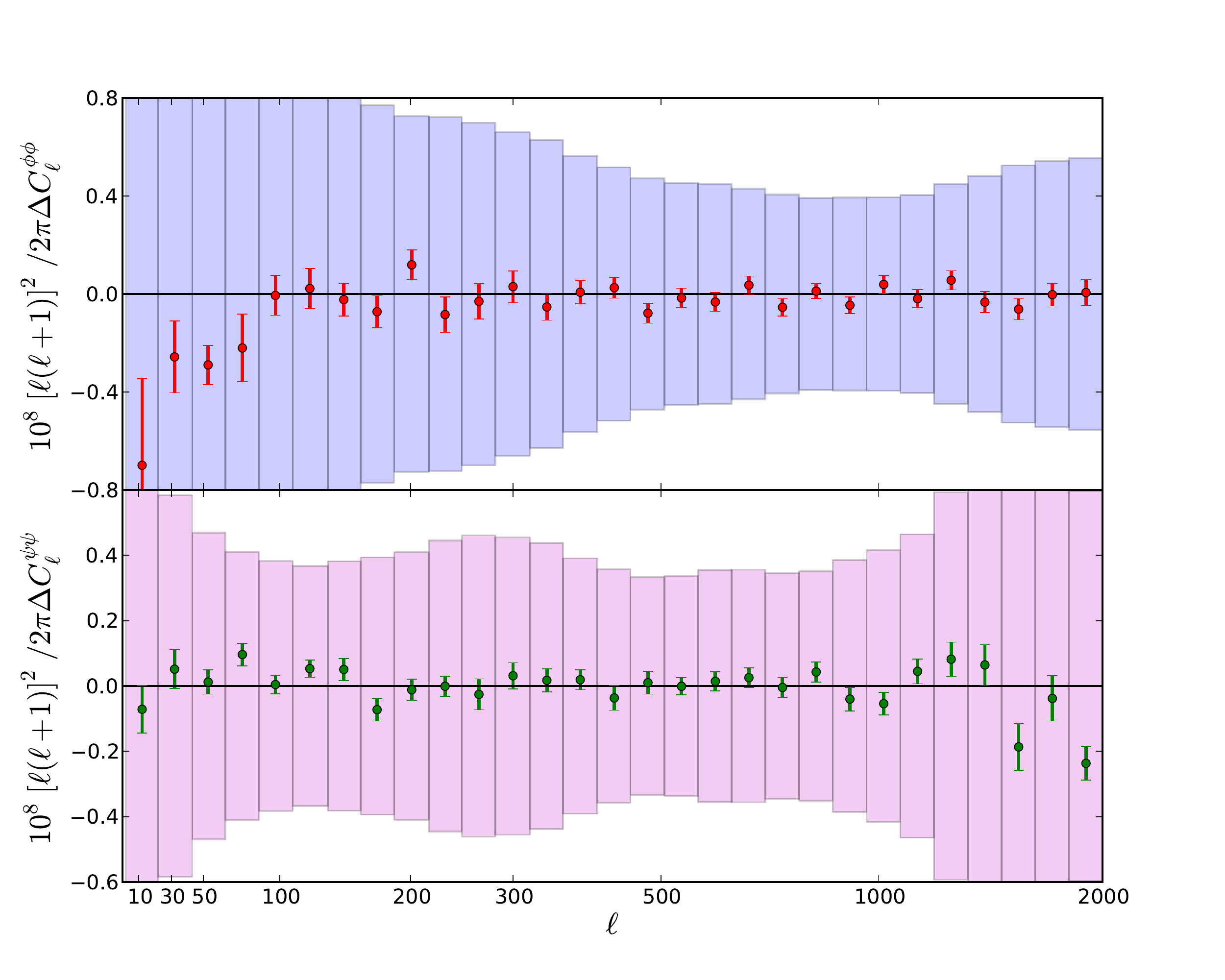}

\caption{Residuals over 100 lensed simulations of the gradient (top) and curl (bottom) modes of the deflection angle in presence of the Galactic mask, apodized over \mbox{$\theta_{\rm{apo}}=10^\circ$}. Histograms represent the error bars for one reconstruction (Eq. \ref{var_mask}).\label{rec_mask} }
\end{center}
\end{figure}

The final reconstruction is presented on Fig. \ref{rec_mask} where we show the mean over 100 residuals of the reconstructed lensing potential power spectrum. In this case, the reconstruction seems less in agreement with the fiducial spectrum at low multipoles, but it should be emphasized that the small bias that can be seen is well with the expected errors bars, whereas the mask-mean field is actually several orders of magnitude higher that the lensing potential at those scales.

\subsection{Normalization}
The rescaling of the power spectrum of the reconstructed map by $1/f_{\rm{gal,4}}$ can be thought as a modification of the normalization factor applied to the unnormalized estimator $\bar{g}_{\ell m}$. Indeed, rescaling the power spectrum by  $1/f_{\rm{gal,4}}$ is equivalent to applying a normalization factor equal to $A^\phi_\ell f_{\rm{gal,4}}^{-1/2}$. We can elaborate on this by estimating the normalization factor to apply to the unnormalized reconstructed potential by computing it by a Monte-Carlo procedure. There are two ways of doing so, either by correlating the lensing potential reconstructed from a lensed simulation with the input lensing potential used to generate the lensed temperature map, or by correlating two lensing reconstructions from two different temperature maps lensed with the same lensing potential. We thus form the following correlations that define the Monte-Carlo normalizations $A^{\phi, MC,2}_\ell$  and $A^{\phi, MC, 4}_\ell$
\be
\frac{1}{N}\sum_{i=1}^N\langle \bar{g}^i_{\ell m} \phi^{th}_{\ell m}\rangle =\frac{C_\ell^{\phi\phi, th}}{A^{\phi, MC,2}_\ell}\label{A2}
\ee

\be
\frac{2}{N(N-1)}\sum_{i=1}^N \sum_{j=i+1}^N \langle \bar{g}^i_{\ell m} \bar{g}^j_{\ell m}\rangle =\frac{C_\ell^{\phi\phi, th}}{(A^{\phi, MC,4}_\ell)^2}\label{A4}, 
\ee
where $\bar{g}^i_{\ell m}$ and $\bar{g}^j_{\ell m}$ are to be understood as reconstructions from different realizations of temperature maps lensed with the same lensing potential. As Eq. (\ref{A2}) brings in the two-point correlation function of the masked temperature and Eq. (\ref{A4}) the four-point correlation function, relations between $A^{\phi, MC,2}_\ell$, $A^{\phi, MC, 4}_\ell$ and $A^{\phi}_\ell$ depend on $f_{\rm{gal},2}$ and $f_{\rm{gal},4}$, where 
\be 
f_{\rm{gal,2}}=\frac{1}{N_{\rm{pix}}} \sum_i w_i^2.
\ee
More precisely, we have
\be
A^{\phi, MC,2}_\ell= \frac{A^{\phi}_\ell}{f_{\rm{gal,2}}}, \mbox{ and } A^{\phi, MC,4}_\ell= \frac{{A^{\phi}_\ell}}{ \sqrt{f_{\rm{gal,4}}}}. 
\ee 

\begin{figure}[htb]
\begin{center}
\includegraphics[width=1.00\columnwidth]{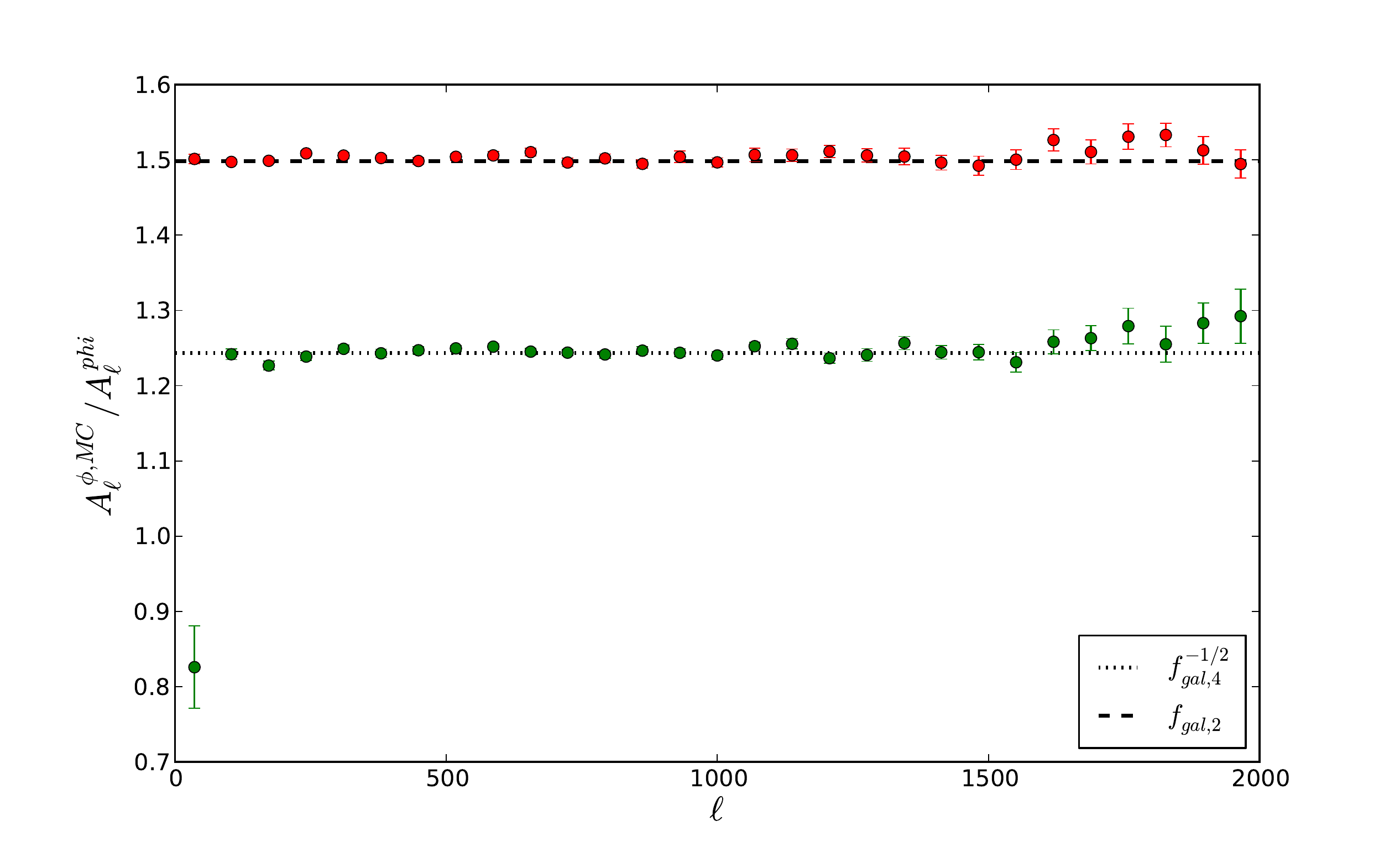}

\caption{\label{normm}Ratio of the normalization computed by Monte-Carlo to the theoretical normalization in the case of a reconstruction with an apodized Galactic mask.}
\end{center}
\end{figure}

Figure \ref{normm} presents the ratio of the normalization computed by Monte-Carlo to the theoretical normalization. Red dots  correspond to $A^{\phi, MC,2}_\ell$ and green dots correspond to $A^{\phi, MC,4}_\ell$. The first bin is not in agreement because of the presence of the mask mean field which yield a non-zero correlation in Eq. (\ref{A4}). Accordingly, when computing the normalization by correlating the reconstructed lensing potential to the input potential, the mask mean field disappears from the correlation. This figure shows a very good agreement between the ratio and the expected constant values, which validates once again the use of an apodized Galactic mask for CMB lensing reconstruction. It should be noted that the two normalizations computed here do not have the same utilization. We have seen previously that $A^{\phi, MC,4}_\ell$ is used to normalize the power spectrum of the reconstructed lensing potential. As $A^{\phi, MC,2}_\ell$ only brings in a single reconstructed potential in its definition, it should be used to normalize correlations between the reconstructed lensing potential and any other field that would have a non-zero correlation with the lending potential but not with the reconstruction noise. A perfect example for such a field would be an external dataset of tracers of the large scale structure.

\section{Point source and galactic mask}

All the ingredients are now in place to reconstruct the lensing potential and its power spectrum when both the point sources and Galactic masks are included. The algorithm is fairly simple. We first inpaint  the regions masked by the point sources mask using the algorithm described in Sec. \ref{sec:cork}. We then apply the apodized Galactic mask on this inpainted map. We chose to use the largest apodization length we considered previously, $\theta_{\rm{apo}}=10^\circ$.
We then generate a set of unlensed simulations  under the lensed temperature power spectrum on which we run the quadratic estimator to estimate the mask mean-field.

Renormalization of the reconstructed lensing potential power spectrum requires some care as some of the masked regions in the Galactic plane do not have any signal and must therefore by rescaled by the Galactic mask renormalization value  $f_{\rm{gal},4}$. However the inpainted regions lack the lensing signal, but possess the full Gaussian structure, so that the Gaussian noise coming from those regions is the theoretical $N_\ell^{(0)}$. The estimator of the lensing potential power spectrum then takes the following form

\be
C_L^{\tilde \phi \tilde\phi, \rm{mask+PS}}= \frac{1}{f_{\rm{PS}}^2}\left[\frac{\langle g_{LM}g^*_{L M}  \rangle }{f_{\rm{gal},4}}- \left(1+\frac{1}{N_{\rm unl}^{\rm MF}} \right)N_L^{(0.g)}\right] - N_L^{(1,g)}\label{rec_clpp_mask_PS},
\ee
and its variance becomes
\beqn
\sigma^2\left(C_\ell^{ \tilde \phi \tilde \phi, \rm{mask+PS}}\right)&=& \frac{f_{\rm{gal},8}}{f^4_{\rm{PS}}f^2_{\rm{gal},4}}\frac{2}{2\ell+1}\\\nonumber
&&\left[ C_\ell^{\phi\phi, \rm{fid}} +\left(1+\frac{1}{N_{\rm unl}^{\rm MF}} \right) N^{(0,g)}_\ell + N^{(1,g)}_\ell   \right]^2
\label{var_maskps}\eeqn

\begin{figure}[htb]
\begin{center}
\includegraphics[width=1.00\columnwidth]{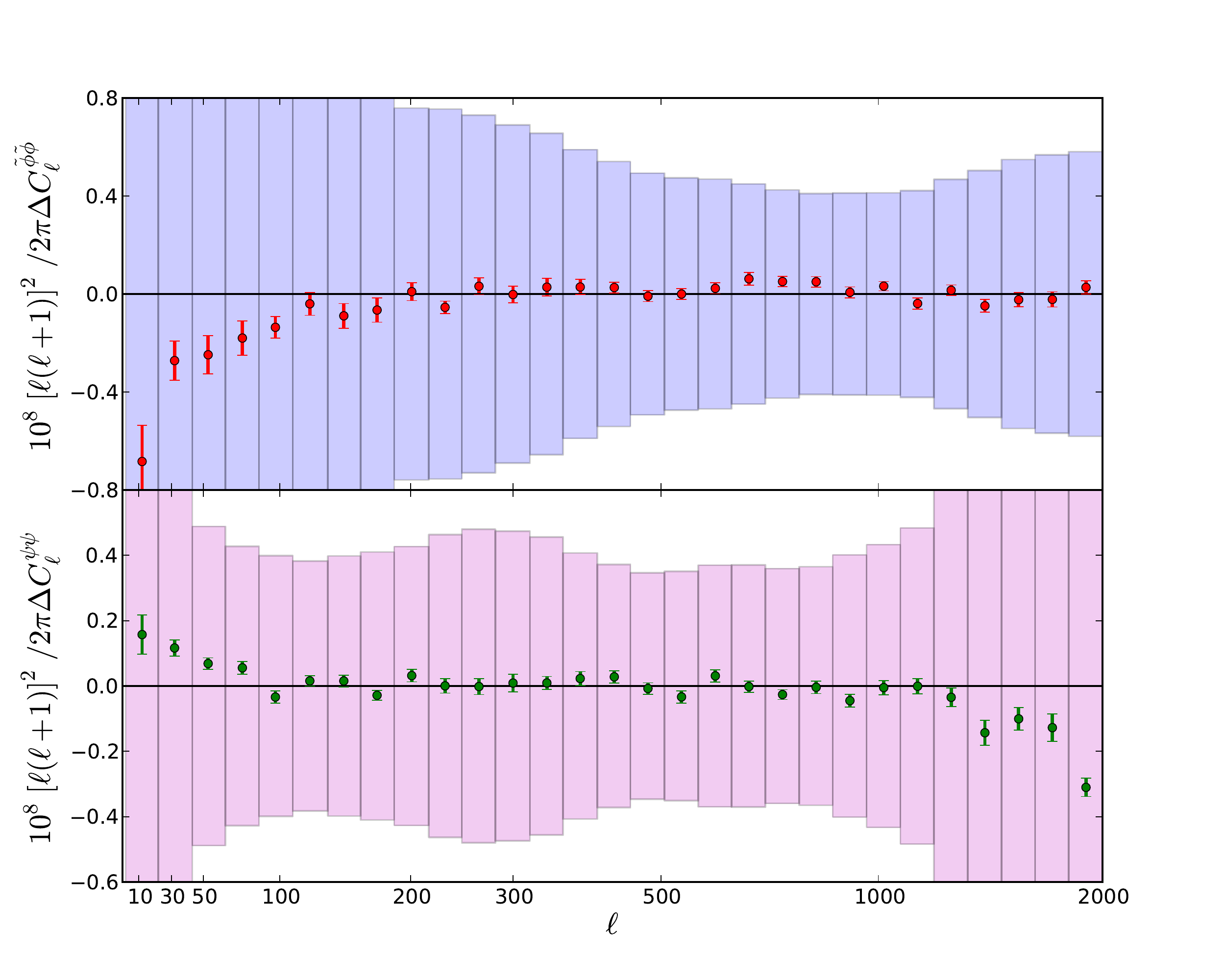}

\caption{Residuals over 300 simulations of the gradient (top) and curl (bottom) modes of the deflection angle in presence of the point sources mask and of the Galactic mask, apodized over \mbox{$\theta_{\rm{apo}}=10^\circ$}. Histograms represent the error bars for one reconstruction (Eq. \ref{var_maskps}).\label{rec_mask_source} }
\end{center}
\end{figure}

The residuals over 300 lensed simulations of the gradient and curl power spectra reconstruction are presented in Fig. \ref{rec_mask_source}. We notice the presence of biases  at low multipoles both for gradient and curl modes and at high multipoles for the curl modes, but their amplitude is small, about 0.1$\sigma$. 
However, the residuals show excellent agreement with zero in the multipole range $100\leq \ell\leq 1000$, for both the gradient and the curl modes of the deflection angle. This multipole range is also where the signal-to-noise ratio is the highest, thus validating the described pipeline.

\section{Discussion}

We have presented a new, simple, and robust pipeline for CMB lensing reconstruction. This pipeline is designed to treat two important issues in a full-sky lensing reconstruction analysis: the presence of point sources that need to be masked and Galactic foregrounds. We treat the first problem by filling in the holes of the temperature map by Gaussian constrained realizations. This operation perfectly restores the Gaussian structure of the original map and inpainted maps can then safely be ingested in the lensing estimator.

The issue of Galactic contamination is treated by simply applying an apodized Galactic mask to the temperature map before the lensing estimation. The presence of the mask creates a bias in the reconstructed potential. However, since this bias is solely related to the mask and does not depend on the actual lensing potential, it can be efficiently estimated from unlensed simulations and then subtracted to the final result. This simple operation is enough to guarantee an unbiased reconstruction of the lensing power spectrum. The contribution of unresolved  point sources to the lensed CMB trispectrum is assumed to be small and is therefore ignored.

By using an isotropic filtering during the reconstruction process, this pipeline remains analytical, in the sense that both the normalization $A^\phi_\ell$ and the Gaussian noise  $N^{0}_\ell$ keep their theoretical expression as they are just properly rescaled by scale-independent factor to account for the missing power due to the presence of masked  region. 

In this article, we have used a simple noise model by considering that the detector noise is white and homogeneous. This will not be the case for a realistic experiment like Planck, where the scanning strategy lead to a non-uniform coverage of the sky. Analytical predictions of the effect of inhomogeneous noise on lensing reconstruction have been investigated in \citet{Hanson09}. The pipeline described in this work can easily be generalized to inhomogeneous noise, by taking into account the noise structure in the unlensed simulations when the mask mean-field is computed. In that case, the estimated mean-field would be the sum of the noise and mask mean-fields. Similarly, any other systematic effects that would create a fake lensing potential can be treated in a similar way, the difficulty being more in the ability to correctly simulate these effects than in the lensing reconstruction in itself.

 \section*{Acknowledgments}
ABL acknowledges fruitful discussions with E. Hivon. ABL is supported by the Leverhulme Trust and STFC. DH acknowledges support from CITA National Fellowship. Part of this work has been initiated with the CMB lensing working group of the Planck Collaboration. We also acknowledge the use of the HEALPix package and the cosmological code CAMB.

\bibliographystyle{aa}
\bibliography{./biblio_lensing2}

\end{document}